\documentstyle[epsfig,11pt]{article}
\def\ltsima{$\; \buildrel < \over \sim \;$}
\def\simlt{\lower.5ex\hbox{\ltsima}}
\def\gtsima{$\; \buildrel > \over \sim \;$}
\def\simgt{\lower.5ex\hbox{\gtsima}}

\begin{document}
\vspace{1.0cm}
{\Large \bf THE BEPPOSAX X-RAY VIEW OF REFLECTION-DOMINATED SEYFERT GALAXIES}

\vspace{1.0cm}

M.Guainazzi$^1$, G.Matt$^2$ and F.Fiore$^3$

\vspace{1.0cm}
$^1${\it Astrophysics Division, Space Science Department of ESA, ESTEC/SA, Postbus 299, NL-2200 AG Noordwijk, The Netherlands}\\
\centerline{mguainaz@astro.estec.esa.nl} \\
$^2${\it Dipartimento di Fisica ``E.Amaldi'', Universit\`a degli Studi ``Roma Tre'', Via della Vasca Navale 84, I-00146 Roma, Italy}\\
$^3${\it Osservatorio Astronomico di Roma, Via dell'Osservatorio, I-00144, Monteporzio Catone, Italy}\\

\vspace{0.5cm}

\section*{ABSTRACT}
We present new results from BeppoSAX observations of reflection-dominated
Seyfert galaxies, and namely: 1) the Compton--thick Seyfert 2
NGC~1068 and Circinus Galaxy; 2) the Seyfert 1 NGC~4051, whose nucleus was
observed on May 1998 to have switched off, leaving only a residual reflection
component as an echo of its past activity. Our main focus in this paper
is on the soft X-ray continuum properties and on the X-ray line spectroscopy.

\section{COMPTON-THICK SEYFERT~2 GALAXIES}

Despite the fact that Seyfert galaxies are classified according mainly to
optical properties, the X-ray behavior of type 1 and 2 objects is
remarkably different,  the main difference laying in
the amount of photoelectric absorption from neutral matter.
Seyfert~2s normally exhibit
column densities ${\rm N_H \simgt 10^{21}}$~cm$^{-2}$ (Awaki et al.
1991; Turner at el. 1997), but in a few cases
no significant column density is observed above the Galactic
contribution. A key to understand this apparent oddity came
from the discovery of broad optical lines in
the polarized optical spectra of several Seyfert~2s (Antonucci \& Miller
1985; Tran 1995; Heisler et al. 1997), which posed
the basis for the so called ``Seyfert unification theories''
(Antonucci 1993). In some ``Compton-thick'' objects,
the interposing matter is optically thick to Compton
scattering, and therefore the impinging photons are down-scattered to
energies where the photoabsorption cross section becomes dominant. In these
cases the primary nuclear continuum is invisible, unless scattered
and/or reflected along our line of sight. X-rays might follow the
same optical path as the optical photons, allowing view of the radiation
produced in the innermost nuclear regions, close to the central black hole.

If this is indeed the case, one might expect that the scattering
plasma (``warm mirror'') adds a wealth
of emission lines from ionized elements, due to 
fluorescence/recombination and/or resonant scattering.
ASCA observations unveiled that the X-ray spectrum of Compton-thick
Seyfert~2s is indeed rich of emission lines, mainly due to He-like stages
of intermediate elements (Matt et al. 1997b; Iwasawa et al. 1997; Turner et al.
1997). However, in all these cases the flat observed continuum
(photon index ${\Gamma < 1}$) and a huge
fluorescent iron line from neutral matter (Equivalent Width ${\rm EW >1}$~keV)
suggest that the intermediate X-ray spectrum is dominated by
Compton reflection from optically thick neutral matter
(Matt et al. 1997b). Moreover, several
Seyfert~2s host intense nuclear starbursts, which are expected to
produce a not negligible contribution to the soft X-ray spectrum (Ptak et al.
1998). The spectra of Compton-thick Seyfert~2s are therefore complex
and the limited ASCA energy bandpass strongly hampers an
accurate and unique deconvolution,
which is essential to evaluate the properties of the emission lines, in
particular the intensity against their proper continuum.

The scientific payload on board BeppoSAX (Boella et al. 1997)
covers the unprecedented wide
energy interval between 0.1 and 200~keV, and hence allows to circumvent
the problem encountered by ASCA. In this section, we will report the results
of the observation of two close and bright Compton-thick galaxies
observed during the AO1 phase: NGC~1068 and Circinus Galaxy.
Both these galaxies have polarized broad optical lines
(Antonucci \& Miller 1985; Oliva et al. 1998) and host
intense nuclear starburst rings  (Scoville 1988; Maiolino et al. 1998).
Details on the data
reduction can be found in Matt et al. (1997a). A more detailed analysis
can be found in Guainazzi et al. (1999). 

\subsection{\underline{Main Observational Results}}

The hard X-ray ({\it i.e.}: $\simgt 4$~keV) spectra of
NGC~1068 and Circinus Galaxy exhibit a substantial amount of
Compton-reflection (Matt et al. 1997a; Matt et al 1999),
which dominates above several
keV. The photon spectral index $\Gamma$
of the primary component is $\simeq$2.1 in NGC~1068
and $\simeq$1.6 in Circinus Galaxy. These values are close to those observed
in Seyfert~1 galaxies (Nandra et al. 1997), as foreseen by the
unified theories.
In Circinus Galaxy a prominent excess in the PDS band ({\it i.e.}: $\simgt
10$~keV) can be most easily explained as due to the transmission
of the same primary continuum through an absorbing screen of $\simeq
4 \times 10^{24}$~cm$^{-2}$. The case of Circinus is particularly intriguing,
since a cutoff in the primary continuum is required (${\rm E_{cutoff} \simeq
60}$~keV). This is the first time that such a cutoff is detected in a
Seyfert~2.

\begin{figure}
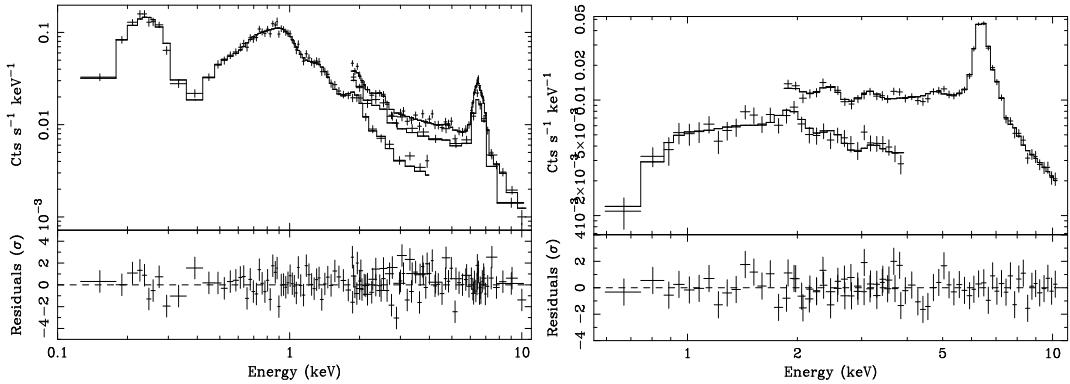

\hspace{1.0cm}
\epsfig{figure=fig1a.ps,width=5cm,height=7cm,angle=-90}
\epsfig{figure=fig2a.ps,width=5cm,height=7cm,angle=-90}
\caption{Spectra and residuals in units of standard
deviations when the best fit model is applied to NGC~1068
({\it left}) and Circinus Galaxy ({\it right}). The best fit model is
composed of: a bare Compton-reflection continuum, a scattering component
and a single temperature optically thin plasma emission (model
MEKAL in XSPEC). Best-fit parameters are reported in Table~2}
\label{fig1}
\end{figure}

\begin{figure}
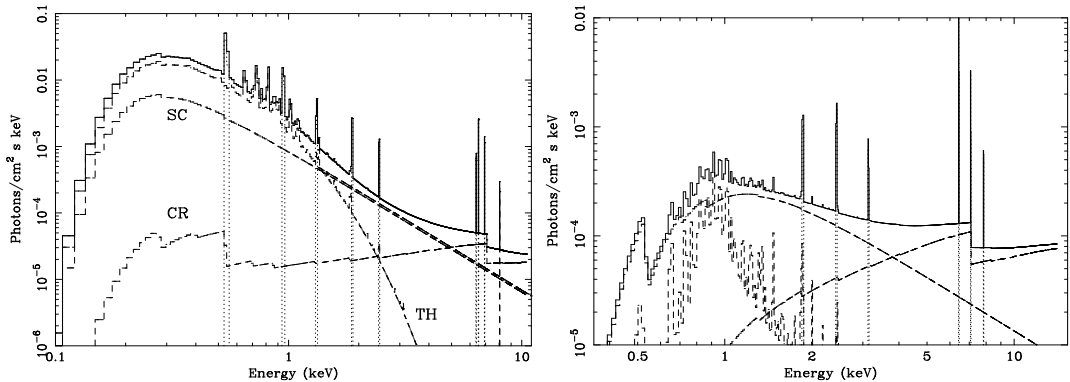

\hspace{1.0cm}
\epsfig{figure=fig1b.ps,width=5cm,height=7cm,angle=-90}
\epsfig{figure=fig2b.ps,width=5cm,height=7cm,angle=-90}
\caption{Best-fit models for NGC~1068
({\it left}) and Circinus Galaxy ({\it right}). The continuum
spectra components are labeled: CR (Compton reflection);
SC (scattering), TH (thermal plasma)}
\label{fig2}
\end{figure}

A prominent soft excess is present in both sources, along with
a rich set of emission lines (see Table~\ref{tab1}). The
\begin{table}
\caption{Emission lines in the BeppoSAX spectra of NGC~1068 and
Circinus Galaxy}
\label{tab1}
\vspace{0.5cm}
\begin{center}
\begin{tabular}{lcccc} \hline \hline
& \multicolumn{2}{c}{NGC~1068} & \multicolumn{2}{c}{Circinus Galaxy} \\ 
Identification & ${\rm E_C}$ & ${\rm EW}$ & ${\rm E_C}$ & ${\rm EW}$ \\ 
& (keV) & (eV) & (keV) & (eV) \\ \hline
O{\sc vii} & $0.54 \pm 0.04$ & $310 \pm 140$$^{\star}$ & ... & ... \\
Ne{\sc ix} & $0.95 \pm 0.03$ & $320 \pm 100$$^{\star}$ & ... & ... \\
Mg{\sc xi} & $1.32 \pm 0.05$ & $170 \pm 70$$^{\star}$ & $1.32$$^{\dag}$ & $<50$$^{\star}$ \\
Si{\sc xiii} & $1.88 \pm 0.03$ & $280 \pm 60$$^{\star}$ & $1.87 \pm^{0.04}_{0.05}$ & $160 \pm^{80}_{60}$$^{\star}$ \\
S{\sc xv} & $2.45 \pm0.04$ & $210 \pm 50$$^{\star}$ & $2.43 \pm^{0.05}_{0.04}$ & $260 \pm 70$$^{\star}$ \\
Ar{\sc xvii} & 3.1$^{\dag}$ & $<5$$^{\star}$ & $3.14 \pm^{0.11}_{0.09}$ & $140 \pm^{60}_{80}$$^{\star}$ \\[0.5cm]
\multicolumn{5}{l}{Iron (nickel?) lines:} \\
``Neutral'' & $6.4$$^{\dag}$ & $1600 \pm 200$$^{\diamond}$ & $6.446 \pm^{0.012}_{0.013}$ & $2850 \pm 130$$^{\diamond}$ \\
``Neutral'' (K$_{\beta}$) & ... & ... & $7.08 \pm 0.06$ & $640 \pm^{100}_{110}$$^{\diamond}$ \\
Fe{\sc xxv} & 6.7$^{\dag}$ & $3000 \pm 600$$^{\star}$ & 6.7$^{\dag}$ & $<70$$^{\star}$ \\
Fe{\sc xxv} (K$_{\beta}$) (or Ni K$_{\alpha}$) & ... & ... & $7.9 \pm^{0.2}_{0.3}$ & $700 \pm 300$$^{\star}$ \\
Fe{\sc xxvi} & 6.96$^{\dag}$ & $1500 \pm^{400}_{500}$$^{\star}$ & ... & ... \\
Fe{\sc xxvi} (K$_{\beta}$) & $8.1 \pm 0.2$ & $600 \pm 300$$^{\star}$ & ... & ... \\ \hline \hline
\end{tabular}
\end{center}
\noindent
\hspace{2.0cm}$^{\dag}$fixed \hspace{1.0cm}
$^{\star}$against the scattering \hspace{1.0cm}
$^{\diamond}$against the Compton reflection
\end{table}
soft excess can be only partly
accounted for by scattering of the primary continuum, and
{\it cannot} be modeled by a multi-temperature
optically thin emission, as suggested by Ueno et al.
(1994)
The most convincing explanation
for the observed continua is the composition of scattering and of a
single temperature optically thin plasma emission. Spectra and best-fit
models are shown in Figure~\ref{fig1} and ~\ref{fig2}, respectively,
while a summary of the best-fit
\vspace{1.0cm}
parameters is reported in Table~\ref{tab2}.

\subsection{\underline{Discussion}}

The continuum in the
objects of our sample can be deconvolved in at least three separate
components (which are likely to be produced in physically and spatially
different regions): a) Compton-reflection from neutral or
mildly ionized matter (and the associated fluorescent iron line);
b) scattering of a Seyfert~1-like primary continuum
(and the associated fluorescent ionized emission lines);
c) optically thin thermal plasma.

The most likely origin for the Compton reflection
is the matter responsible for the
complete obscuration of the nuclear continuum
({\it e.g.}:
the molecular torus envisaged by the unification theories).
High-resolution imaging in the optical (Malkan et al. 1997)
and IR (Granato et al. 1997; Maiolino et al. 1998) showed that the
distribution of matter on the scale of tens of pc is very
complex, with dusty lanes or bars protruding towards the center, which
might as well contribute to the X-ray extinction.

In the case of Circinus Galaxy, the 2.8~keV EW iron line, coupled with
the relatively low inclination angle derived from the hard X-ray
analysis (Matt et al. 1999), implies an iron
overabundance by at least a factor of 3 (Matt et al. 1996a; Matt et al.
1997b). In NGC~1068 the situation is far more complex. If we assume the
iron line deconvolution in Table~\ref{tab1}, the EW of the neutral component
is $\simlt 1.6$~keV, suggesting a high inclination. This is in agreement
with the idea that the line comes from the same region as the water maser
(${\rm \imath \simgt 82^{\circ}}$, Greenhill et al. 1996).

The warm scattered continuum is described as a fainter replica
of the intrinsic nuclear power-law. If the spectral indices of the primary
and scattered continuum are left independently free to vary in the fit,
no significant difference is found (in NGC~1068: ${\rm \Delta \Gamma \equiv
\Gamma_{primary} - \Gamma_{scattered} = 0.3\pm^{0.4}_{0.5}}$; in
Circinus Galaxy, ${\rm \Delta \Gamma = 0.1 \pm^{1.6}_{1.0}}$).
\begin{table}
\caption{Continuum parameters for NGC~1068 and Circinus Galaxy best-fit models}
\label{tab2}
\vspace{0.5cm}
\begin{center}
\begin{tabular}{lcc} \hline \hline
& NGC~1068 & Circinus Galaxy \\
${\rm N_H}$ ($10^{20}$~cm$^{-2}$) & $3.1 \pm 0.3$ & $80 \pm^{60}_{50}$ \\
$\Gamma$ & $2.13 \pm 0.17$ & $1.6 \pm^{0.2}_{0.3}$ \\
${\rm F_{Compton}}$$^{\ddag}$ ($10^{-12}$~erg~cm$^{-2}$~s$^{-1}$) & 1.9 & 6.3 \\
${\rm F_{scattering}}$$^{\ddag}$ ($10^{-12}$~erg~cm$^{-2}$~s$^{-1}$) & 2.2 & 3.0 \\
${\rm L_{th}}$ ($10^{41}$~erg~s$^{-1}$)$^{\dag}$ & 3.7 & 0.14 \\
${\rm kT_{th}}$ & $440 \pm 50$ & $500 \pm^{700}_{300}$ \\
${\rm A_Z}$ (\%) & $2.8 \pm 1.6$ & $23 \pm^{465}_{22}$ \\
$\chi^2_{\nu}$ & 1.02 & 0.86 \\ \hline \hline
\end{tabular}
\end{center}
\noindent
\hspace{3.5cm}$^{\ddag}$in the 2--10~keV band \hspace{1.0cm}
$^{\dag}$in the 0.5--4.5~keV band
\end{table}
The bulk of the ionized lines observed in the
X-ray spectra of both targets is likely to originate as resonant
scattering or fluorescence in the same medium.
In Circinus Galaxy the lack of ionized iron emission lines is consistent
with a single-zone scatterer with intermediate ionization parameter.
On the other hand, in NGC~1068
the wide range
of elements involved (and, consequently, of ionization stages required
to support the existence of different elements in He- or H-like stages)
implies that the reflector need not to be one-zone or homogeneous. The
most likely configuration is a continuum distribution of ionization
parameters ${\rm \xi}$
in the whole range $10^2$--$10^4$ (Netzer et al. 1997). Following
Matt et al. (1996a), we used the ratio of the Compton-reflected (which
depends mainly on the inclination angle) with the warm scattered flux
(which depends almost entirely on the optical depth of the scattering
material) to derive an estimate of the column density
of the warm mirror: $\simlt$ a few $10^{21}$~cm$^{-2}$ in NGC~1068
and a few $10^{22}$~cm$^{-2}$ in Circinus Galaxy.

A ${\rm N_{H,warm} \sim 10^{22}}$~cm$^{-2}$ scatterer should
also imprint absorption feature in the soft X-ray spectrum. The addition
of photoionization absorption edges, however,
does not yield any improvement in the quality of the fit. The upper limits
on the optical depths of O{\sc vii} and O{\sc viii} photoionization edges
are 0.35 and 0.25 in NGC~1068, respectively. The same limits are 2.7 and
1.5 in the
low Galactic latitude (and therefore highly absorbed) Circinus Galaxy.
Assuming that the scattering plasma is in the typical conditions of the
``warm absorbers'' observed in Seyfert~1 galaxies (Reynolds 1997; George et 
al. 1998), this corresponds to hydrogen equivalent column densities
of ${\rm N_{H,warm} \simlt 5 \times 10^{21}}$~cm$^{-2}$ and
${\rm N_{H,warm} \simlt 4 \times 10^{22}}$~cm$^{-2}$, broadly in agreement
with the above diagnostics. More conclusive statements
have to be deferred to
future high-resolution soft X-ray spectroscopical data.

We tentatively associate the thermal component with the contribution
of the intense nuclear starburst. David et al. (1992) derived an empirical
correlation between the FIR and the X-ray luminosity of starbursts.
If we apply their scale law, the expected X-ray luminosities are
$\simeq 10^{41}$~erg~s$^{-1}$ and $\simeq 2.1 \times 10^{40}$~erg~s$^{-1}$
for NGC~1068 and Circinus Galaxy,
respectively, broadly consistent with the observed ones, given also the
admittedly strong uncertainties on the correlation parameters. This
strengthens the case for our identification. A breakthrough
in this field will be provided by space-resolved spectroscopy on
arc seconds scale, which the forthcoming {\it Chandra}
high-resolution detectors will
allow.

\section{SOFT X-RAYS IN THE OFF STATE OF THE SEYFERT~1 GALAXY NGC~4051}

\begin{figure}
\begin{center}
\epsfig{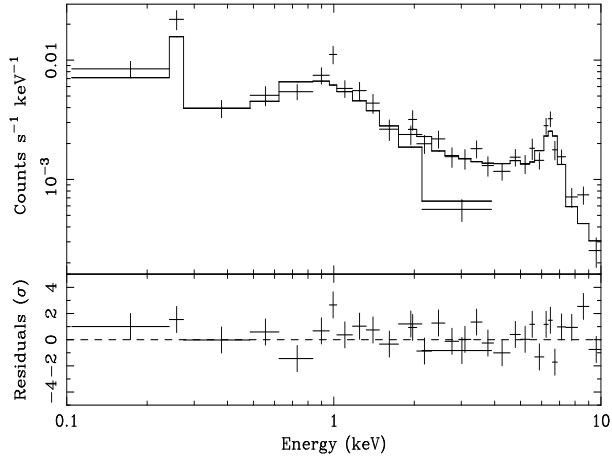}
\caption{Spectra ({\it upper panels}) and residuals in units
of standard deviations ({\it lower panels}), when the broadband
BeppoSAX spectrum is fitted with a simple power--law + ``bare
reflection'' model}
\end{center}
\label{fig5}
\end{figure}

The strongly X-ray variable (Lawrence et al. 1987; Matusoka et al. 1990;
Guainazzi et al. 1996; Uttley et al. 1998) and low-luminosity
Seyfert~1 galaxy NGC~4051
was observed by BeppoSAX on May 1998 in an ultra-low and stable state,
which was best explained assuming that the active nucleus had switched off,
leaving only a residual reflection component visible (Guainazzi et al.
1998). A similar flux decrease was observed only in the ``fading'' Seyfert~2
NGC~2992 (Weaver et al. 1992).
A prominent soft excess above the extrapolation of the hard X--ray best--fit
is evident below 4~keV. This soft component cannot be unambiguously
modeled. A power-law, an optically thin
plasma, or a double blackbody provide comparably good fits,
as shown in Table~\ref{tab3} and Figure~\ref{fig5},
while a scattering scenario {\it alone} is ruled out.
The 0.5--2~keV (0.1--2~keV) {\it observed} flux
\begin{table}[hbt]
\begin{footnotesize}
\begin{tabular}{lccccc} \hline \hline
Soft excess model & ${\rm N_H}$ & ${\rm \Gamma_{hard}}$ & ${\rm \Gamma_{soft}}$ or ${\rm kT^1_{bb}}$ or ${\rm kT_{mekal}}$ & ${\rm N_{soft}/N_{hard}}$ or ${\rm kT^2_{bb}}$ or ${\rm Z_{mekal}}$ & $\chi^2/$~dof \\
& ($10^{20}$~cm$^{-2}$) & & (keV) &  (keV) or (\%) & \\ \hline
Power--law & $3.4\pm^{1.2}_{1.0}$ & $1.75\pm^{0.18}_{0.15}$ & $3.0\pm^{0.2}_{0.3}$ & $13\pm^{13}_{6}$ & 116.2/93 \\
{\verb!mekal!} & 1.18$^{\ddag}$ & $1.92\pm^{0.19}_{0.13}$ & $0.75\pm^{0.09}_{0.10}$ & $< 1.5$ & 111.6/92 \\
Double blackbody & $< 2.0$ & $1.98\pm^{0.19}_{0.16}$ & $0.31 \pm^{0.05}_{0.04}$ & $0.095\pm^{0.014}_{0.012}$ & 115.9/81 \\ \hline \hline
\end{tabular}
\\
\noindent
$^{\ddag}$~unconstrained
\caption{Best--fit parameters for the broadband NGC4051 Beppo-SAX spectrum.
Different models in column~1 correspond to different
description of the soft excess. ${\rm \Gamma_{hard}}$ is the photon
index of the nuclear continuum, which gives rise to the
Compton reflection component dominating above 4~keV}
\label{tab3}
\end{footnotesize}
\end{table}
is $\simeq 7.4 \ (12.1) \times 10^{-13}$~erg~cm$^{-2}$~s$^{-1}$,
corresponding to an {\it unabsorbed}
rest frame luminosity of $1.7 (7.0) \times 10^{40}$~erg~s$^{-1}$.
There is no evidence of absorption edges and/or emission lines,
the 90\% upper limit on the optical depth of {\sc
O vii} and {\sc O viii} photoabsorption edges being 0.34 and 0.13,
respectively. The steepness of
the spectrum and the lack of any ``warm absorber'' imprinting,
strongly point against a nuclear origin of the soft X--rays.
The intrinsic 0.1--2~keV luminosity inferred by the present
data is high but
not uncommon for a ``normal" galaxy (Fabbiano 1989).
It is therefore possible that
the dimming of the nucleus has left the galactic components 
(supernovae winds, hot halos) as the bulk
of the observed soft X--rays.
Alternatively, the observed soft X-rays
could have the same origin as the extended
(spatial scale $\sim 100$~pc) emission observed by the
ROSAT HRI in NGC~4151
(Morse et al. 1995), which coincides with the optical narrow line
emitting clouds. It was interpreted as thermal emission from
a hot (${\rm T \sim 10^7}$~K) and low density (${\rm n_e < 1}$~cm$^{-3}$)
gas, in pressure equilibrium with the Narrow Line Region clouds.
Again, space-resolved spectroscopy would be invaluable under this
respect, if the NGC~4051 will be so kind to let herself to be caught again
in a ultra-dim state.

\section{REFERENCES}
\vspace{-5mm}
\begin{itemize}
\setlength{\itemindent}{-8mm}
\setlength{\itemsep}{-1mm}

\item[] Antonucci R., ARA\&A, {\bf 31}, 473 (1993)

\item[] Antonucci R. \& Miller J.S., ApJ, {\bf 297}, 621 (1985)

\item[] Awaki H., Koyama K., Inoue H., Halpern J.P., {\bf 43}, 195 (1991)

\item[] Boella G., Butler R., Perola G.C., A\&AS, {\bf 112}, 299 (1997)

\item[] David L.P., Jones C., Forman W., ApJ, {\bf 388}, 82 (1992)

\item[] Fabbiano G., ARA\&A, {\bf 27}, 87 (1989)

\item[] George I.M., et al., ApJS, {\bf 114}, 73 (1998)
 
\item[] Granato G.L., Danese L., Franceschini A., ApJ, {\bf 486}, 147 (1997)

\item[] Greenhill L.J., et al., ApJ, {\bf 472}, L21 (1996)

\item[] Guainazzi M., et al., MNRAS, 301, L1 (1998)

\item[] Guainazzi M., et al., MNRAS, submitted (1999)

\item[] Guainazzi M., Mihara T., Otani C., Matsuoka M., PASJ, {\bf 48}, 781 (1996)

\item[] Heisler C.A., Lumsen S.L., Bailey J.A., Nature, {\bf 385}, 700 (1997)

\item[] Iwasawa K., Fabian A.C., Matt G., MNRAS, {\bf 289}, 443 (1997)

\item[] Lawrence A., Watson M.G., Pounds K.A., Elvis M., Nature, {\bf 325}, 694 (1987)

\item[] Maiolino R., et al., ApJ, {\bf 493}, 650 (1998)

\item[] Matsuoka M., Piro L., Yamauchi M., Murakami M., ApJ, {\bf 361}, 440 (1990)

\item[] Matt G., Brandt W.N., Fabian A.C., MNRAS, {\bf 280}, 823 (1996a)

\item[] Matt G., et al., A\&A, {\bf 325}, L13 (1997a)

\item[] Matt G., et al., A\&A, {\bf 341}, L39 (1999)

\item[] Matt G., et al., MNRAS, {\bf 281}, 69 (1996b)

\item[] Matt G., Fabian A.C., Reynolds C.S., MNRAS, {\bf 289}, 175 (1997b)

\item[] Morse J.A., Wilson A.S., Elvis M., Weaver K.A., ApJ. {\bf 439}, 121 (1995)

\item[] Nandra K., George I.M., Mushotzky R.F., Turner T.J., Yaqoob T., ApJ, {\bf 477}, 602 (1997)

\item[] Netzer H., Turner T.J., George I.M., ApJ, {\bf 488}, 694 (1997)

\item[] Oliva E., Marconi A., Cimatti A., Di Serego Alighieri S., A\&A, {\bf 329}, L21 (1998)

\item[] Ptak A., Serlemitsos P., Yaqoob T., Mushotzky R., ApJS, in press (astroph/9808159, 1999)

\item[] Reynolds C.S., MNRAS, {\bf 286}, 513  (1997)

\item[] Scoville N.Z., ApJ, {\bf 327}, 61 (1988)

\item[] Tran H.D., ApJ, {\bf 440}, 565 (1995)

\item[] Turner T.J., George I.M., Nandra K., Mushotzky R.F., ApJS, {\bf 113}, 23 (1997)

\item[] Ueno S., et al., PASJ, {\bf 46}, L71 (1994)

\item[] Uttley P., McHardy I., Papadakis I.E., Cagnoni I., Fruscione A., Nucl. Phys. B., {\bf 69/1-3}, 490 (1998)

\item[] Weaver K.A., et al., ApJ, {\bf 458}, 160 (1996)

\end{itemize}

\end{document}